\documentclass[sigconf]{acmart}
\usepackage{multirow}
\usepackage{xcolor}
\usepackage{colortbl}
\usepackage{tcolorbox}
\usepackage{listings}
\usepackage{makecell}
\usepackage{graphicx}

\makeatletter

\newcommand{\Rmnum}[1]{\expandafter\@slowromancap\romannumeral #1@}
\makeatother

\definecolor{codegreen}{rgb}{0,0.6,0}
\definecolor{codegray}{rgb}{0.5,0.5,0.5}
\definecolor{codepurple}{rgb}{0.58,0,0.82}
\definecolor{backcolour}{rgb}{0.95,0.95,0.92}

\lstdefinestyle{mystyle}{
    backgroundcolor=\color{backcolour},   
    commentstyle=\color{codegreen},
    keywordstyle=\color{magenta},
    numberstyle=\tiny\color{codegray},
    stringstyle=\color{codepurple},
    basicstyle=\ttfamily\footnotesize,
    breakatwhitespace=false,         
    breaklines=true,                 
    captionpos=b,                    
    keepspaces=true,                 
    numbers=left,                    
    numbersep=5pt,                  
    showspaces=false,                
    showstringspaces=false,
    showtabs=false,                  
    tabsize=2
}

\usepackage[]{collab}
\collabAuthor{yc}{blue}{YC'Note}
\collabAuthor{yt}{red}{YT'Note}
\collabAuthor{py}{orange}{PY'Note}

\AtBeginDocument{%
  \providecommand\BibTeX{{%
    \normalfont B\kern-0.5em{\scshape i\kern-0.25em b}\kern-0.8em\TeX}}}

\copyrightyear{2024} 
\acmYear{2024} 
\setcopyright{acmlicensed}\acmConference[LLM4Code '24]{2024 International Workshop on Large Language Models for Code}{April 20, 2024}{Lisbon, Portugal}
\acmBooktitle{2024 International Workshop on Large Language Models for Code (LLM4Code '24), April 20, 2024, Lisbon, Portugal}
\acmDOI{10.1145/3643795.3648392}
\acmISBN{979-8-4007-0579-3/24/04}

\newcommand{\fixedwidth}[1]{{\ttfamily \small #1}}
\newcommand{\name}[0]{IDECoder}

\begin{document}

\title{Enhancing LLM-Based Coding Tools through Native Integration of IDE-Derived Static Context}

\author{Yichen Li}
\email{ycli21@cse.cuhk.edu.hk}
\affiliation{%
  \institution{The Chinese University of Hong Kong}
  \department{Department of Computer Science and Engineering}
  \city{Hong Kong SAR}
  \country{China}
}

\author{Yun Peng}
\email{ypeng@cse.cuhk.edu.hk}
\affiliation{%
  \institution{The Chinese University of Hong Kong}
  \department{Department of Computer Science and Engineering}
  \city{Hong Kong SAR}
  \country{China}
}

\author{Yintong Huo}
\authornote{** indicateds the corresponding author.}
\authornotemark[1]
\email{ythuo@cse.cuhk.edu.hk}
\affiliation{%
  \institution{The Chinese University of Hong Kong}
  \department{Department of Computer Science and Engineering}
  \city{Hong Kong SAR}
  \country{China}
}

\author{Michael R. Lyu}
\email{lyu@cse.cuhk.edu.hk}
\affiliation{%
  \institution{The Chinese University of Hong Kong}
  \department{Department of Computer Science and Engineering}
  \city{Hong Kong SAR}
  \country{China}
}





\begin{abstract}
Large Language Models (LLMs) have achieved remarkable success in code completion, as evidenced by their essential roles in developing code assistant services such as Copilot. Being trained on in-file contexts, current LLMs are quite effective in completing code for single source files. However, it is challenging for them to conduct repository-level code completion for large software projects that require cross-file information. Existing research on LLM-based repository-level code completion identifies and integrates cross-file contexts, but it suffers from low accuracy and limited context length of LLMs. In this paper, we argue that Integrated Development Environments (IDEs) can provide direct, accurate and real-time cross-file information for repository-level code completion. We propose \name, a practical framework that leverages IDE native static contexts for cross-context construction and diagnosis results for self-refinement. IDECoder utilizes the rich cross-context information available in IDEs to enhance the capabilities of LLMs of repository-level code completion. We conducted preliminary experiments to validate the performance of IDECoder and observed that this synergy represents a promising trend for future exploration.

\end{abstract}


\maketitle

\section{Introduction}

Inspired by the great success of large language models (LLMs) on natural language processing (NLP), code LLMs, including CodeX~\cite{codex} and StarCoder~\cite{li2023starcoder}, are proposed and obtain promising performance in code intelligence tasks such as code completion. The code LLMs broadly advance the development of code assistant services like Copilot~\cite{copilot_doc} and CodeWhisperer~\cite{codewhisperer}. By providing real-time, context-based code suggestions, code assistant services significantly improve developer productivity.

 \begin{figure}[tbp]
  \centering
     \includegraphics[width=\columnwidth]{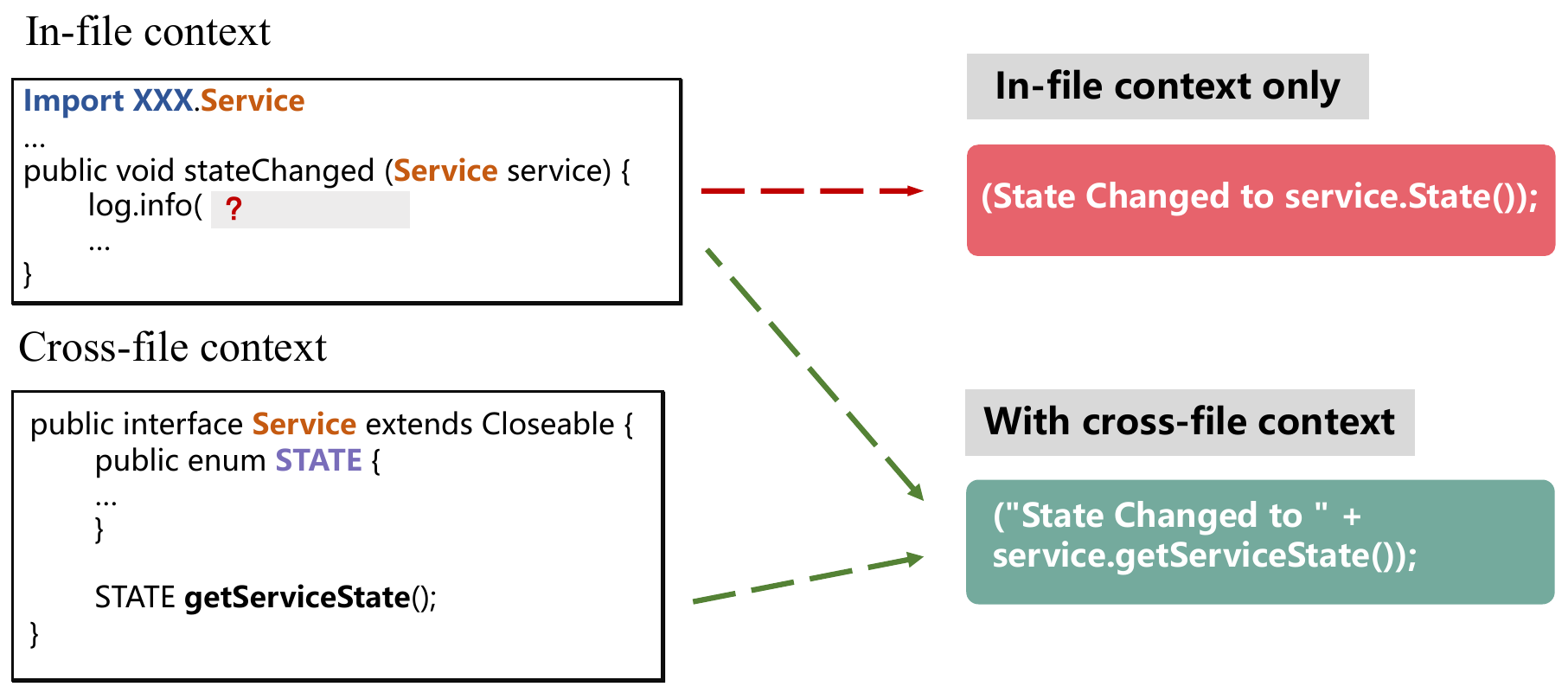}
     \vspace{-0.2in}
     \caption{An example of the completed code from StarCoder. StarCoder fails to complete a Java program correctly as the in-file context does not provide sufficient type information of \textit{Service}. To get the service state, the model needs to know the member function \textit{getServiceState} of \textit{Service}.}
\vspace{-0.1in}
     \label{fig:case}
 \end{figure}
 
Similar to general LLMs, code LLMs are trained on in-file contexts~\cite{li2023starcoder, codex, codegeex} and are not aware of software project architectures. Without cross-file information such as imported methods and external classes, they struggle to understand the entire architecture of a software project and obtain significantly low performance in repository-level (repo-level) code completion~\cite{ding2023crosscodeeval}. Fig.~\ref{fig:case} presents a failed case of StarCoder~\cite{li2023starcoder} in repo-level code completion. Even with a large knowledge base obtained in the pre-training stage, Code LLM fails to complete the logging statement since the in-file context does not provide sufficient type information of variable \textit{Service}, which is defined in another file and may not appear in the training data. Therefore, it is essential to include cross-file information, which cannot be obtained in the internal knowledge base, for code LLMs in repo-level code completion.

To include cross-file context information, researchers have proposed various repo-level code generation frameworks~\cite{zhang2023repocoder,ding2022cocomic,bairi2023codeplan,deepseek-coder, shrivastava2023repofusion}. Most adopt two major steps: identifying the appropriate cross-file context and logically integrating cross-file contexts. To identify cross-file contexts, they employ techniques such as identifier name matching~\cite{ding2022cocomic} and semantics retrieval~\cite{gao2023constructing, zhang2023repocoder}. For fusing cross-file contexts, they utilize methods like all-import~\cite{deepseek-coder} and encoding-based fusion~\cite{shrivastava2023repofusion}. These approaches can supplement the in-file contexts with external knowledge. However, they still suffer from unsatisfactory performance due to various factors, such as inaccurate context identification caused by complicated language features and limited context length of LLMs.

Modern Integrated Development Environments (IDEs)~\cite{pylance,idea} natively offering robust code navigation, suggestion and static diagnosis capabilities, are widely used by developers in software development. IDEs can manage the entire project structure and provide practical cross-file context information, such as class hierarchies, function signatures, and variable types in real time. Despite the impressive performance of current LLM-powered coding tools, they do not make use of the rich context information available within IDEs and thus struggle to handle repo-level code completion.

In this work, we argue that LLMs and LLM-powered coding tools should address code completion tasks by jointly considering in-file and cross-file contexts, leveraging the native static information and static diagnosis result (warnings, errors) directly provided by Integrated Development Environments (IDEs)~\cite{wei2023copiloting} during development. We propose \name, a novel and practical LLM-based code completion framework powered by IDE native static contexts. Our vision is to integrate native static information as static contexts and diagnosis results as feedback for self-refinement.  This framework allows LLMs to utilize the wealth of cross-file information in IDEs, ultimately enhancing their capabilities to complete code that is aware of the overall project structure and dependencies. We conduct preliminary experiments of IDECoder which demonstrate the promise of IDE native code LLMs as a promising and practical direction of exploration. By integrating static contexts and diagnosis results, our framework paves the way for more sophisticated and context-aware code completion, ultimately benefiting developers in their software development.

\section{Repo-level Code Completion: Challenges}

 \begin{figure*}[tbp]
  \centering
     \includegraphics[width=1.9\columnwidth]{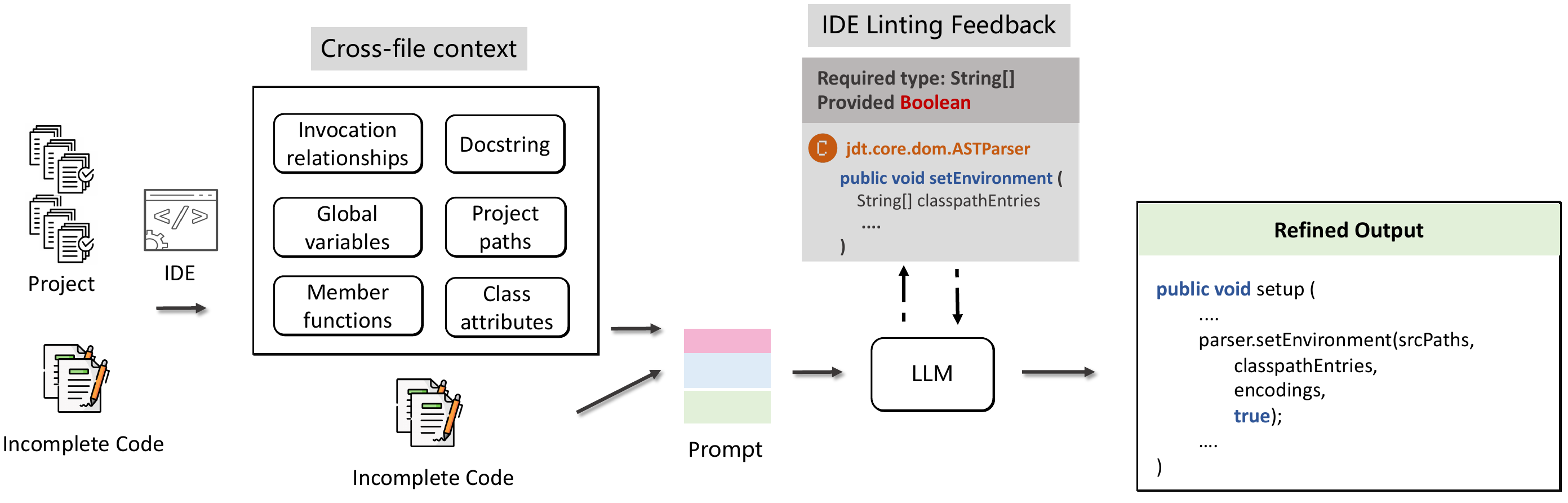}
     \vspace{-0.1in}
     \caption{The overall framework of IDECoder.}
\vspace{-0.1in}
     \label{fig:me}
 \end{figure*}

Repo-level code completion aims to complete code using the cross-file broader contexts. However, it faces difficulties in accessing useful and relevant information dispersed across files, such as imported classes and their member functions from different modules, global variables, and external classes with unknown type details. This process involves two primary steps: cross-file context identification and context fusion. In the following subsections, we introduce the challenges for the two steps.

\subsection{Identification of Cross-file Contexts}

When working with a code snippet and associated files in a repository, it is crucial to identify the most relevant and useful cross-file contexts, such as member functions of utilized variables, for code-based LLMs. This ensures accurate reference and prevents confusion for LLMs. The primary challenges in identifying cross-file contexts are maintaining \textit{Accuracy} and ensuring \textit{Relevance}.

\textit{Accuracy.} Current retrieval-based methods~\cite{zhang2023repocoder,ding2022cocomic,lu2022reacc} of identifying related references based on identity names and semantic similarity can hardly handle complicated program language features, such as inheritance, polymorphism, and complex namespaces, and thus leading to wrong identifications. Accurately identifying the cross-file contexts necessitates a more refined approach considering the specific language features. In contrast to Retrieval-Augmented Generation (RAG)~\cite{lewis2020retrieval} tasks based on semantic similarity, fuzzy matching is generally not applicable for code-related tasks under the soundness requirement. Therefore, there is a need for developing static analysis-enhanced retrieval approaches.

\textit{Relevance.} Accurate reference and definition identification are not enough for building cross-file contexts. It is also crucial to identify relevant contexts that reflect the developer's intentions, such as code comments~\cite{ding2022cocomic}. This helps LLMs understand the relationships across the repository, such as the expected method invocation sequences, the functional role of the current file within the entire project, and the module dependencies. Existing approaches~\cite{gao2023constructing,ding2023logentext} provide the code LLMs with similar code snippets via in-context learning to teach code-based LLMs how to program in the current case. However, LLMs do not actually understand the developer's intention in the current project and the role of the current source file since examples come from other software projects and are not relevant to the current project.

\subsection{Fusion of Cross-file Contexts}

Given the collected cross-file contexts, it is essential to organize them in a way that code-based LLMs can better understand. Simply concatenating in-file and cross-file contexts is suboptimal for two reasons. First, it is impractical to include all source codes of identified import classes, due to
the limited context length of LLMs. Second, overlong contexts can also hurt the performance of current code-based LLMs~\cite{ding2022cocomic} and lead to excessive costs~\cite{ChatGPT}. Current encoding-based methods~\cite{ding2022cocomic,shrivastava2023repofusion} simply fuse these contexts together, but they still need LLMs to be tuned for the new data format. Besides, approaches that rely solely on in-context learning~\cite{gao2023constructing} by providing input and output examples may exacerbate the drawbacks of similarity-based retrieval methods.

Furthermore, different elements in the contexts should have different importance. For example, local dependencies in the current projects are more important than popular third-party packages (i.e., Numpy~\cite{numpy}), as the later ones may already exist in the internal knowledge base of LLMs.

\section{Methodology}

The IDECoder aims to integrate static information as static contexts and diagnosis results in LLM-based code completion. As illustrate in Fig.~\ref{fig:me}, IDECoder takes the target incomplete code and its corresponding project as input in IDE. By leveraging the analysis capabilities of IDEs, IDECoder accurately identifies cross-file contexts and extracts relevant information for cross-file context identification. Next, IDECoder models and organizes the identified cross-file contexts as inputs for the LLMs. It employs a chain-of-thought~\cite{peng2023generative} methodology to model this information sequentially, enabling the LLM to generate more contextually relevant code completions. Finally, IDECoder refines the generated code using the diagnostic output from the IDE's linting service, which ensures the quality and correctness of the generated code.

To effectively integrate cross-file context information into the LLM-based code completion process, our methodology focuses on three key phases: \textit{cross-file context identification}, \textit{cross-file context fusion}, and \textit{linting-based code refinement}.

\subsection{Cross-file Context Identification}

To address the challenges of cross-file context identification, we take advantage of the native capabilities of IDEs to pinpoint above mentioned cross-file contexts.

In terms of \textit{Accuracy}, IDEs provide various features such as abstract syntax tree (AST) construction, symbol table~\cite{power2000symbol} creation, reference indexing, and code element localization. By utilizing these features, IDEs can parse code, identify code elements, references and further relationships, and look up class attributes. With language-specific static analysis, IDEs can accurately identify cross-file contexts. This allows IDECoder to overcome the limitations of current retrieval-based methods~\cite{ding2022cocomic, zhang2023repocoder,zan2023private} by considering language-specific features like inheritance, polymorphism, and complex namespaces.

For \textit{Relevance}, IDEs can assist in identifying the most pertinent cross-file contexts that reflect the developers' intentions, such as docstrings of imported methods, method invocation relationships, method functionalities, and module dependencies. By understanding the relationships of different code elements within the repository, IDEs can provide valuable contextual information to LLMs, enabling them to generate code that aligns with the developers' intentions and project structure.

\subsection{Cross-file Context Fusion}
Given the identified cross-file contexts, IDECoder then organizes them as inputs for LLMs. Unlike previous methods such as encoding~\cite{ding2022cocomic,shrivastava2023repofusion}, all import~\cite{deepseek-coder}, reasoning~\cite{bairi2023codeplan} that use the complete code of methods in cross-file contexts, IDECoder utilizes docstrings, as well as method and class signatures with detailed type information. Docstrings indicate the developers' intentions of building the methods, while method and class signatures reflect the functionalities. Using them instead of the entire code could largely reduce the context length without losing valuable information.

To collect docstrings and method and class signatures, IDECoder first analyzes the imported packages and modules in the repository to differentiate between widely used third-party libraries and user-defined classes and methods. IDECoder incorporates different strategies for handling third-party libraries and user-defined classes and methods. For \textbf{third-party} libraries, such as NumPy~\cite{numpy} in Python, IDECoder briefly includes version information of the package to ensure API as the code and usage patterns of such libraries may already exist in the knowledge base of LLMs obtained in the pre-training stage~\cite{huang2023not}. For \textbf{user-defined} classes and methods, IDECoder extracts their essential properties and relationships. It then collects docstrings and method and class signatures with detailed type information acquired from class and method definitions. By leveraging such information, IDECoder can create a concise and informative representation of the cross-file context, enabling LLMs to make accurate and contextually relevant code completions.

Given the collected docstrings and method and class signatures, IDECoder does not directly list them as the inputs. Instead, it employs a chain-of-thought methodology~\cite{peng2023generative} to model information, sequentially illustrating each piece of information. This approach introduces the acquired context in a top-down manner, ranging from the functional role to specific type details.  This mechanism allows IDECoder to consider the sequence of code elements, their relationships, and the developers' intentions when generating code completions. By incorporating this information, IDECoder can generate more contextually relevant and coherent code snippets that adhere to the project's structure and design principles. Once the chain-of-thought prompts are constructed, they are sent to the LLMs for initial results.

\subsection{Linting-based Code Refinement}

Given the completed code generated by LLMs, IDECoder further conducts linting-based code refinement to ensure the quality and correctness of the generated code.

Current IDEs utilize a range of as-you-type static code inspection tools, such as Pylance~\cite{pylance} and ESLint~\cite{idea}, to identify and rectify incorrect code within a project before compilation. Once IDECoder receives completion results from LLMs, the IDE's as-you-type linting service is activated to detect potential warnings, type errors, and method usage issues.

IDECoder employs a two-step process to ensure the quality of completed code. First, it utilizes the diagnostic output from the linting service to identify issues~\cite{wei2023copiloting} in the generated code. Then, it resends the identified issues and their previous context to the LLMs for correction and refinement. This process enhances the generated code's quality and allows IDECoder to self-improve and rectify variable usage, ensuring the generated code's syntactical correctness. In the case of unimported used methods in completed code, the import management process is optimized to efficiently maintain import packages with the suggestions of linting, overcoming the limitations of plain text style completion. 


After refining the code based on the linting feedback, \name presents the corrected code to the user. Linting-based code refinement offers a more lightweight, real-time, and unobtrusive user experience compared to conversational program repair~\cite{xia2023conversational} using execution feedback, as it does not necessitate program execution.

\section{Preliminary Evaluation}\label{sec:eva}

\subsection{Preliminary Implementation} 
We access the static information by hooking into Pylance~\cite{pylance} as the \textbf{proof-of-concept} to showcase the effectiveness of IDECoder in this vision paper. Pylance is an extension that works alongside Python in Visual Studio Code~\cite{vscode} to provide Python language services for programmers. All contexts for prompt construction come from hooking into events of the Pylance plugin. For some strongly-coupled information that cannot be hooked, we \textbf{manually} collect and send them to LLMs in our experiments.

We set GPT-3.5 (the fixed version, \textit{gpt-3.5-turbo-0301}) as the backbone model. Besides, We adjusted the \textit{temperature} to 0 to ensure the model consistent generation, thereby ensuring reproducibility.

\subsection{Evaluation Setup}

\noindent
\textbf{Datasets.} Owing to the lack of repo-level code completion datasets, we emulate the data collection pipeline from the code completion dataset that has cross-file contextual information \fixedwidth{CROSSCODEEVAL}~\cite{ding2023crosscodeeval}. Other cross-file datasets~\cite{ding2023crosscodeeval,ding2022cocomic} are not applicable because they do not allow customized prompts as inputs for  by offering the fixed cross-file prompts.
We randomly sample 10 Python repositories from \fixedwidth{CROSSCODEEVAL} to conduct the function body completion task~\cite{zhang2023repocoder} (functions less than 15 lines) and use the GPT-3.5 backbone model, as it can comprehend complex prompt strategies.

\noindent
\textbf{Metrics.} In line with previous works~\cite{gao2023constructing, chakraborty2022natgen, ren2020codebleu}, we evaluate the performance of code completion with three metrics including Exact Match (EM), CodeBLEU (CB)~\cite{ren2020codebleu}, and Syntax Match (SM). EM evaluates the extent to which the model-generated code is identical to the target code. CB calculates the similarity of code snippets by considering syntax and semantic information, such as data flow and Abstract Syntax Tree (AST). SM quantifies the match of subtrees within the code.

\subsection{Evaluation Results}

As a preliminary step, we implement three methods as baselines for comparison: an in-file completion method, an all-import method~\cite{deepseek-coder}, and a basic RAG~\cite{zhang2023repocoder} method following the previous work~\cite{zhang2023repocoder}.

\begin{table}[h]
\centering

\setlength{\tabcolsep}{1.2mm}
\renewcommand{\arraystretch}{1.2}

\caption{Performance comparison on the function body completion dataset. Numbers are shown in percentage (\%).}
\vspace{-0.1in}
\label{tab:result}

\begin{tabular}{l|ccc}

\toprule
Metrics     & Exact Match & CodeBLEU & Syntax Match  \\
\midrule

In-file   & 7.37 & 27.84& 44.17 \\
All-import & 6.71  &  30.53   & 46.24 \\
RAG & 9.77 &   31.65  & 48.92 \\
\name & \textbf{10.46}  &  \textbf{34.16} & \textbf{50.73}  \\
\bottomrule
\end{tabular}

\label{time_split}
\end{table}

Based on the evaluation results presented in Table. ~\ref{tab:result}, IDECoder consistently outperforms all baseline methods in code completion tasks, highlighting its effectiveness. The performance of the other frameworks in these tasks underlines the existing limitations in cross-file code completion. \textit{Note that} the current implementation of IDECoder is constrained by the closed-source Pylance plugin, which restricts full access to the aforementioned contexts. In future work, if IDE vendors develop customizable native coding tools that allow for more extensive access to cross-file contexts, the performance of the IDECoder framework could potentially surpass its current capabilities.
\section{Discussion and Future Plans}
Code LLMs are becoming increasingly influential and gradually integrating into the daily lives of developers, benefits millions of users. 
LLMs natively integrated with Integrated Development Environments (IDEs) are expected to be even more powerful and natural. 
The strong capabilities of IDEs in retrieving information and profiling project structures largely enhance current large-scale code models.
Our preliminary experiment (seen in Sec. ~\ref{sec:eva}) results support this claim, demonstrating a promising research direction of learning cross-file contexts.

In the future, we plan to implement a more mature version of IDECoder by developing the IDE plugin, which can support user-defined backbone LLMs. Additionally, we will extend IDECoder to support a broader range of code-related tasks. As the core idea of extending cross-file contexts from IDE-provided static information for code intelligence tasks has shown success in this paper, it can be generalized to other code-related tasks, such as repo-level automated program repair (APR) and IDE-assisted automated debugging.

\section{ACKNOWLEDGEMENT}

The work described in this paper was supported by the Research Grants Council of the Hong Kong Special Administrative Region, China (No. CUHK 14206921 of the General Research Fund). We thank all reviewers for their valuable comments.

\bibliographystyle{ACM-Reference-Format}
\bibliography{sample-base}



\end{document}